\documentclass{cjaa}                    

\usepackage{graphicx}                   
\input{epsf.sty}                        
\input{psfig.sty}                       

\begin{document}
\title{Status and Future Prospects for $\gamma$-ray Polarimetry}

   \author{M. L. McConnell
      \inst{1}\mailto{}
   \and P. F. Bloser
      \inst{1}
       }
   \offprints{M. L. McConnell}                   

   \institute{Space Science Center, University of New Hampshire, Durham, NH  03824\\
             \email{Mark.McConnell@unh.edu}
          }

   \date{Received~~2001 month day; accepted~~2001~~month day}

   \abstract{
The recent detection of linear polarization from GRB120206 has piqued the interest of the community in this relatively unexplored avenue of research.  Here, we review the current status and prospects for GRB polarimetry at hard X-ray and soft $\gamma$-ray energies.   After reviewing the most recent results, we present a brief survey of current and planned experiments that are capable of making GRB polarization measurements in the energy range between 30 keV and 30 MeV.
   }

   \authorrunning{M. L. McConnell \ \&  P. F. Bloser}            
   \titlerunning{Prospects for $\gamma$-Ray Polarimetry }  


   \maketitle
%
%

\section{Introduction}

For many years, polarimetry has been used as a useful tool throughout much of the electromagnetic spectrum.  At high photon energies, however, astronomers have been slow to adopt polarimetry as a standard tool.   This has been both because of the experimental difficulty in making such a measurement and because the levels of polarization were expected to be quite low.  Even at lower X-ray energies (1--10 keV), where source fluxes are considerably greater, the community has been slow to embrace the potential value of polarimetry. All this may have changed, however, with the recent detection of $\gamma$-ray polarization from GRB021206.  This result, although controversial, has served to invigorate interest in this area.  This paper provides a brief overview of the experimental status and future prospects of polarimetry at energies above $\sim30$ keV.

\section{Science Background}

Many emission processes that can generate $\gamma$-ray photons can also result in the linear polarization of those photons (e.g., Lei, Dean and Hills, 1997). The level of polarization, however, may depend on the precise emission geometry. In addition, the energy-dependence of the polarization can provide clues to the emission mechanisms that may be operating. Polarization measurements therefore have the potential to tell us something about both the mechanisms and source geometries responsible for the observed emissions. 
At least four emission mechanisms can lead to linearly polarized emissions at hard X-ray energies: 1) the various types of magnetobremsstrahlung radiations (cyclotron, synchrotron and curvature radiation) are all potential sources of linearly polarized emission, depending on the magnetic field configuration; 2) electron-proton bremsstrahlung emission can produce levels of linear polarization of up to 80\%; 
3) initially unpolarized photons can be polarized by Compton scattering, the very same process by which we produce polarized photons in the laboratory; and 4) magnetic photon splitting can lead to polarization levels of up to 30\%. 
Below we outline some specific scenarios for producing linearly polarized hard X-ray and $\gamma$-ray emissions.

\subsection{Gamma-Ray Bursts}

In recent years, largely as a result of the observation of several X-ray, optical and radio afterglows, there has developed a growing consensus that classical $\gamma$-ray bursts (GRBs) are at cosmological distances.  Such great distances imply that a typical GRB releases $10^{51} - 10^{53}$ ergs or more within a time span of several seconds. The general picture that has emerged is one that describes the GRB phenomenom in terms of a relativistic fireball model (e.g., Piran 1999; Hurley, Sari  \&  Djorgovski 2003).  According to this scheme, the $\gamma$-rays are emitted when an ultra-relativistic energy flow is converted into radiation.  The nature of the ``inner engine'' that releases this energy flow is still a subject of speculation.  Once released, however, it is believed that the prompt gamma-ray emission is produced by internal shocks within the outgoing flow.  Internal shocks are generated when one expanding shell overtakes another, slower-moving shell. An external shock arises when the outgoing flow runs into the local interstellar medium (ISM).  The longer wavelength (soft X-ray, optical, radio) afterglow emission is believed to arise from the external shock. 

Many different models have been proposed for the inner engine.  A general characteristic of most models is that they result in the accretion of a relatively massive transient accretion disk ($\sim0.5 M_{\odot}$) onto a black hole. One class of models involves collisions (or mergers) between neutron stars (NS) and/or black holes (BH) in binary systems (e.g., Eichler et al. 1989). The energy released during the merger itself is $\sim5 \times 10^{53}$ ergs (Clark  \&  Eardley 1977), a value comparable to the apparent energy release of cosmological GRBs.  The estimated rate of such mergers is comparable to the observed rate of GRBs (e.g., Piran, 1992), further supporting the idea that such mergers result in GRBs. A second class of models, the so-called collapsar models, are closely related to supernovae. The creation of a collapsar has been described as both a ``failed SN Ib'' (Woosley 1993) or a ``hypernova'' (Paczynski 1998a).  Support for these models comes, in part, from the apparent association of several GRBs with supernovae (e.g., GRB 030329; Stanek et al. 2003) and on the possible association of GRBs with star-forming regions (Paczynski 1998b). 

Another common feature of many GRB models is that the energy release takes the form of jets that are directed along the rotation axis of the system. Several indirect arguments have been used to argue that such jets are required to explain the observations. For example, a break in the afterglow light curve from a relatively shallow power-law decay to a steeper exponential decay can be interpreted as a result of a beam that is laterally expanding with time (e.g., Rhoads 1999; Panaitescu  \&  Kumar 2003). Since the energy budget of a given GRB depends heavily on assumptions about the extent to which the flow is jet-like, determining the reality and nature of jets in GRBs is becoming an important goal of future observations. 

The observation of optical polarization (up to $\sim10$\%) in several GRB afterglows (e.g., Covino et al. 1999; Bersier et al. 2003; Covino et al. 2003) has provided direct evidence for geometrical beaming of emission related to the external shocks.  Several models predict levels of polarization as high as 10 or 20\%, depending on the angle between the observer and the jet axis (e.g., Gruzinov 1999; Gruzinov  \&  Waxman 1999; Sari 1999; Ghisellini  \&  Lazzati 1999).  These optical studies, however, probe only the external shock region. In the context of the canonical fireball model, measurements of the hard X-ray polarization during the prompt phase of the GRB promise to provide a similar probe of the internal shock region.  Since the outgoing flow at the internal shock is expected to be more tightly collimated than the flow at the external shock (resulting from a continuous spreading of the jet as it progresses outward through the fireball), one can expect a somewhat higher level of hard X-ray polarization (assuming that it arises from synchrotron emission) during the prompt phase of the GRB.  Even higher levels of polarization might be expected if the emission results from inverse Compton, rather than synchrotron, emission.  For example, the model of Shaviv and Dar (1995) proposed that the $\gamma$-radiation is generated by Compton scattering off relativistic electrons in the (inner) transient jet, not unlike models proposed for AGN (e.g., Skibo, Dermer  \&  Kinzer 1994). Significant levels of polarization of the hard X-ray emission ($>20$\%) are predicted for all bursts with durations between 1 and 30 seconds.  Even moderate sensitivity to polarization (20-30\%) will help to constrain some of the models.  The recent claim of a very high level of polarization (80\%) at energies $>300$ keV (Coburn  \&  Boggs 2003) underscores the potential for hard X-ray polarization measurements of GRBs. 

The low energy part of the GRB spectrum, which behaves as a power-law, suggests an origin due to synchrotron emission from relativistic electrons, but a simple synchrotron model cannot explain the entire spectrum (Piran 1999). Models for the GRB spectrum typically involve both synchrotron and/or inverse Compton emission (e.g., Shaviv  \&  Dar 1995; Pilla  \&  Loeb 1998). Both mechanisms can lead to significant polarization, but with different energy-dependences.  The uncertainty with regards to the origin of the hard X-ray spectrum is a problem that can also be attacked, at a fundamental level, using polarization measurements. Both inverse Compton and synchrotron emission have been proposed as viable emission mechanisms for GRBs.  The spectral signatures of these two processes can be very similar, so it is very difficult to determine the responsible mechanism on the basis of spectral measurements alone.  Energy-dependent polarization measurements, in principle, offer a possible solution.  The degree of linear polarization of synchrotron emission, unlike inverse Compton emission, is independent of energy.  This energy-dependence of the polarization (or lack thereof) may therefore be exploited as a means of identifying the emission mechanism responsible for the hard X-ray and $\gamma$-ray emission.  

It is widely recognized that the soft $\gamma$-ray repeaters (SGR) represent a different class of phe-nomena than the classical GRBs.  SGRs are short duration, soft-spectrum bursts with super-Eddington luminosities. The bulk of the emission is seen at energies below 100 keV.  A total of four, perhaps five, such sources have now been identified (Hurley 2000).  The prevailing view is that SGR outbursts involve emission from the vicinity of magnetars, neutron stars with magnetic fields in excess of $10^{14}$ G, with the energy release triggered by massive neutron star crustquakes (e.g., Duncan and Thompson, 1992). Baring (1995) suggested that the softness of the events can be attributed to photon splitting  in the extremely intense magnetic fields.  The photon splitting process degrades the high energy $\gamma$-ray photons to hard X-ray energies (Baring 1993). One by-product of photon splitting is that the reprocessed photons would exhibit a polarization level of ~25\% (Baring 1995).  Polarization measurements of the 50--300 keV could therefore provide a test of the importance of photon splitting in SGRs.

\subsection{Solar Flares}

Studies of solar flare $\gamma$-ray line data from the SMM Gamma Ray Spectrometer (GRS) suggest that protons and $\alpha$-particles are likely being accelerated in a rather broad angular distribution (Share  \&  Murphy, 1997; Share et al., 2002).  There is no reason to expect, however, that electrons are being accelerated in a similar fashion. 

Efforts have been made to study electron beaming in solar flares using statistical measurements of the hard X-ray directivity and using stereoscopic observations (as reviewed in McConnell et al. 2002b).  Statistical studies are based on a large sample of solar flares that may not provide sufficient insight into singular events.  Stereoscopic observations are prone to cross-calibration issues.  These  difficulties suggest the need for a technique that can measure time-dependent anisotropies for individual flares using a single instrument.  Polarization is a diagnostic that can meet these requirements.

Many models of nonthermal (e.g., thick target) hard X-ray production predict a clear and significant polarization signal, with polarization levels $>10$\% (Brown, 1972; Langer  \&  Petrosian, 1977; Bai  \&  Ramaty, 1978; Emslie  \&  Vlahos, 1980; Leach  \&  Petroisian, 1983; Zharkova, Brown,  \&  Syniavskii, 1995; Charikov, Guzmna,  \&  Kudryavtsev, 1996).  The precise level of polarization depends on both energy and viewing angle. Some fraction of the observed 20--100 keV hard X-ray flux will be flux that is backscattered from the solar photosphere, the precise magnitude of which will depend, in part, on the polarization of the primary flux.  The reflected component will, in turn, influence the degree of polarization of the observed flux, since even if the electrons are accelerated isotropically, backscattering will introduce polarization fractions of a few percent at energies below 100 keV (e.g., Langer  \&  Petrosian, 1977; Bai  \&  Ramaty, 1978).  Even thermal models of the hard X-ray source predict a finite polarization of order a few percent, due to the anisotropy in the electron distribution function caused by a thermal conductive flux out of the emitting region into the cooler surroundings (Emslie  \&  Brown, 1980). The thermal component, with its rather low polarization, tends to dominate the emission from all flares at energies below about 25 keV.  At these energies, it therefore becomes difficult to distinguish the non-thermal component, with its intrinsic directivity signature, from the thermal component.  This has led to the argument that polarization measurements can best be performed at higher energies (Chanan, Emslie,  \&  Novick, 1988).

These predictions, while clearly testable, could be criticized on the grounds that the modeling assumptions they contain may be overly simplified.  For example, each model to date assumes a single, simple magnetic field structure.  It could be argued that any real flare, particularly one sufficiently large to produce a signal of sufficient strength to enable a polarization measurement, will in all probability contain a mix of structures that would average out any polarization signal present (see also Hudson, Hurford,  \&  Brown 2003).  However, hard X-ray imaging observations in the impulsive phase generally show a fairly simple geometry, consisting of two footpoint sources and perhaps a loop-top source (e.g., Sakao et al., 1992; Masuda et al., 1995). These observations suggest that simple magnetic structures are responsible for the energetic emissions and give support to the possibility that a statistically significant polarization signal could be produced in a large event.

The first measurements of X-ray polarization from solar flares (at energies of $\sim15$ keV) were made by Soviet experimenters using polarimeters aboard the Intercosmos satellites.  In their initial study, Tindo et al. (1970) reported an average polarization for three small X-ray flares of P = 40($\pm$20)\%. This study was followed by an analysis of three flares in October and November of 1970 (Tindo et al., 1972a, 1972b) that showed polarizations of approximately 20\% during the hard impulsive phase.  These reports were met with considerable skepticism, on the grounds that they did not adequately allow for detector cross-calibration issues and limited photon statistics (Brown, McClymont,  \&  McLean 1974). Subsequent observations with an instrument on the OSO-7 satellite seemed to confirm the existence and magnitudes of the polarizations ($\sim$10\%), but these data were compromised by in-flight gain shifts (Nakada, Neupert,  \&  Thomas, 1974).  In a later study using a polarimeter on Intercosmos 11, Tindo et al. (1976) measured polarizations of only a few percent at $\sim$15 keV for two flares in July 1974. This small but finite polarization is consistent with the predictions for purely thermal emission that contains an admixture of polarized backscattered radiation (Bai  \&  Ramaty, 1978). A small polarimeter was flown on an early shuttle flight (STS-3) and made measurements of eight C- and M-class flares in the 5--20 keV energy range.  Upper limits in the range of 2.5\% to 12.7\% were measured, although contamination of the Li scattering material invalidated the pre-flight calibration (Tramiel, Chanan,  \&  Novick 1984).  Most recently, Bogomolov et al. (2005) presented evidence for 20--100 keV polarization levels in excess of 75\% for a solar flare that took place on 29-Oct-2003.

\subsection{Other Sources}

Several other hard X-ray sources may also represent interesting targets for polarization studies (e.g., M{\'e}sz{\'a}ros et al. 1988; Kallman 2004).   Of particular interest will be observations of the Crab Nebula.  Observations indicate a polarization fraction that increases with energy, from 8.1\% at optical wavelengths (Smith et al. 1988) to 19\% at soft X-ray energies (Weisskopf et al. 1978).  The consistency in the polarization angle in these data suggests that a single mechanism, most likely synchrotron radiation, is responsible for the emission.  A similar measurement at higher energies would help determine the extent to which the high-energy emission is related to that at longer wavelengths.

Accreting black hole sources (both AGN and stellar mass black holes) also present an opportunity for useful polarization studies.  Data from OSO-8 provided evidence for small levels of polarization (2-5\%) at low energies ($<10$ keV) from Cyg X-1 (Long et al. 1980), but these observations have never been confirmed.  Sunyaev and Titarchuk (1985) have calculated the degree of linear polarization expected from low-energy photons scattering off hot electrons in an accretion disk.  The degree of polarization depends on the angle of emission with respect to the disk and the optical depth of the emission region.  For optically thin disks, polarization levels as high as 30-60\% are possible.  For structured accretion disks, energy-dependent polarization studies may permit us to probe the details of that structure.  Beamed radiation from accreting black hole sources is also a possible source of polarization (for some of the same reasons that we expect polarization from GRBs).  Beams in AGN, for example, are highly polarized at optical and radio wavelengths, most likely due to synchrotron emission.  Skibo et al. (1994) modeled the hard X-ray emission from Cen A in terms of beamed radiation and predicted high levels of polarization ($\sim60$\%) up to $\sim300$ keV.  Similar models could be considered for many of the so-called ``X-ray novae'' (especially those that exhibit jet features). Some of these sources can occasionally reach intensities that are comparable to the Crab, at which time they would make excellent candidates for polarization studies.

\section{Recent results from RHESSI}

Although originally designed as a hard X-ray solar imager, the Ramaty High Energy Solar Spectroscopic Imager (RHESSI; Lin et al. 2002) has proven itself to be a valuable polarimeter.  Two recent results demonstrate how RHESSI can do polarimetry utilizing two different techniques.  Both techniques make use of RHESSI's 9-element Ge spectrometer array (Figure 1; Smith et al. 2002).  The Ge detectors are segmented, with both a front and rear active volume.  Low energy photons (below about 100 keV) can reach a rear segment of a Ge detector only indirectly, by scattering. For polarization measurements at low energies (20 -- 100 keV), a small block of passive Be (strategically located within the Ge array) is used to scatter photons into the rear segments of adjacent Ge detectors (McConnell et al. 2002b).  Low energy photons from the Sun have a direct path to the Be and have a high probability of Compton scattering into a rear segment of a Ge detector.  The polarization of a transient event (such as a solar flare) can be determined by a careful analysis of the counting rates in the Ge detectors that are closest to the Be scattering block.  This mode is limited to a small FoV ($\sim1^{\circ}$) by the collimation of the Be scattering element through the front of the telescope assembly.   In principle, the capability of RHESSI to simultaneously image the hard X-ray emission represents a major advantage over previous efforts to measure hard X-ray polarization, in that photospherically backscattered photons may be directly imaged by RHESSI as a constraint on the contribution of such backscattered photons to the primary signal.  A preliminary result from this low energy polarimetry mode has indicated a 20--40 keV polarization of $\sim18\%$ from the solar flare of 23-July-2002 (Figure 2; McConnell et al. 2003).

\begin{figure}
	\begin{minipage}[t]{.40\linewidth}
	\centering
	\includegraphics[width=2.00in]{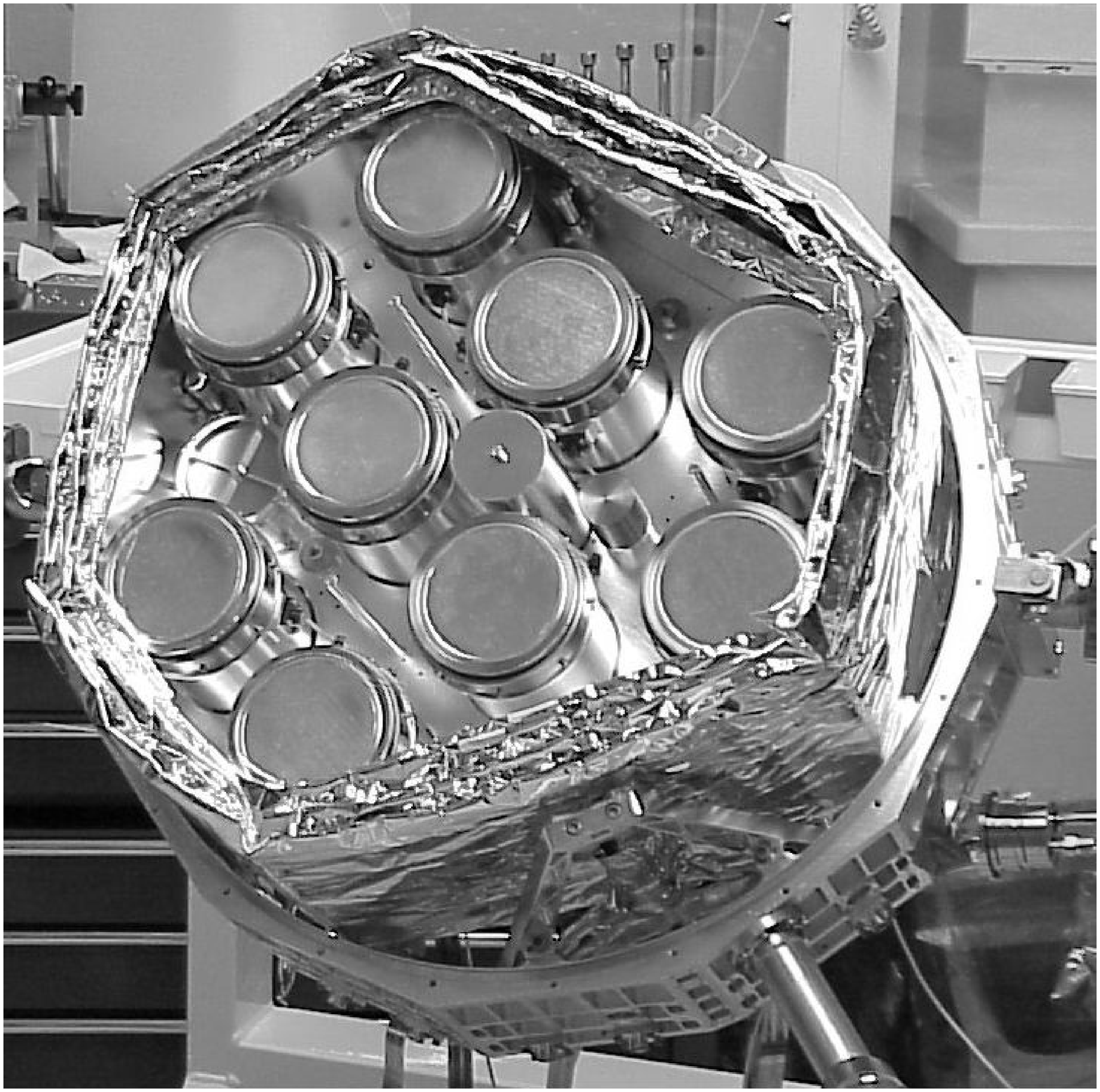}
	\caption{The RHESSI spectrometer array showing the array of nine large Ge detectors.  Near the center of the array is a large LN$_2$ tank used for pre-flight testing.  Just to the right of the LN$_2$ tank is the small Be scattering block used for polarization studies.}
	\label{fig_sim}
	\end{minipage}\hfill
	\begin{minipage}[t]{.55\linewidth}
	\centering
	\includegraphics[width=3.25in]{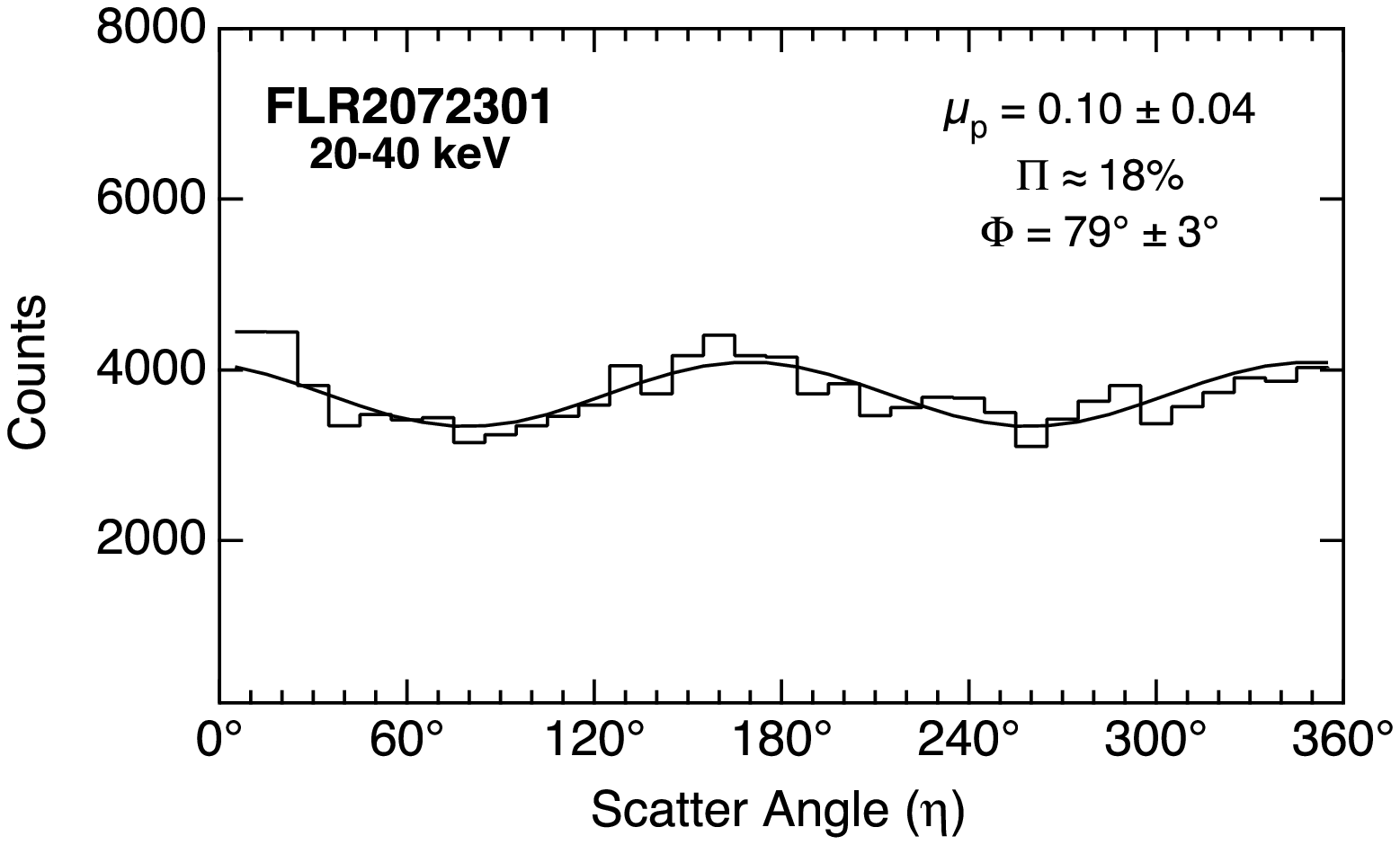}
	\caption{The measured polarization signal as measured by RHESSI in the 20--40 keV energy band for the solar flare of 23-July-2002.}
	\label{fig_sim}
	\end{minipage}
\end{figure}

At higher energies, scattering events between the Ge detectors within the spectrometer array can be used to measure polarization.  The lack of significant amounts of shielding surrounding the Ge array means that this mode is sensitive to events over a much larger area of the sky. In both cases, the rotation of the RHESSI spacecraft (required for imaging with RHESSI's rotation modulation collimators) greatly facilitates effective polarization measurements by reducing systematic uncertainties and providing a more uniform sampling in the azimuthal direction.  This high energy polarimetry mode of RHESSI has been used to derive a result for polarization from GRB 021206 that indicates polarization at a level of $80(\pm20)\%$ in the 150 keV -- 2 MeV energy range (Coburn and Boggs 2003) .   This result, however, has not been confirmed in two independent studies of the same data (Rutledge and Fox 2004; Wigger et al. 2004).

\section{Future Prospects for $\gamma$-Ray Polarimetry}

At these energies, there are three physical processes that can be exploited to measure linear polarization.  These are:  the photoelectric effect, Compton scattering (and its low-energy equivalent, Thomson scattering), and electron-positron pair production.  In each case, the byproducts of the initial photon interaction (photoelectron, scattered photon, or electron-positron pair) have angular distributions that go as $\cos^{2}{\eta}$, where $\eta$ is the angle between the electric field vector of the incident photon and the byproducts' momentum vector.  A measurement of the angular distribution of these secondaries provides a measure of not only the direction but also the magnitude of the linear polarization of the incident flux.  The phase of the distribution is directly related to the direction of the incident polarization.  The amplitude of the modulation in the angular distribution is directly related to the magnitude of the incident polarization. Much of the technical challenge for experimentalists arises from the difficulty in measuring these distributions.

The use of polarimetry in X-ray and $\gamma$-ray astronomy has, to date, been largely limited to energies below $\sim30$ keV (e.g., Angel et al. 1969; Tindo et al. 1970; Novick 1975;  Weisskopf et al. 1990; McConnell et al. 2003b).  Several other experiments have been or are currently being developed for use at these low energies (e.g., Kaaret et al. 1994; Costa et al. 2001; Bellazini et al. 2003; Black et al. 2003).  Here, we concentrate on reviewing those efforts to study polarization at somewhat higher energies (above 30 keV).

\subsection{Low Energy Gamma Rays (30-300 keV)}

Polarimetry in the 30--300 energy band requires low-Z scattering elements (coupled with high-Z photon absorbers) for achieving the best result.  Unfortunately, instruments that operate in this energy band are usually not constructed using position sensitive {\it low-Z} material, but rather they are designed with {\it high-Z} materials to maximize photon absorption.  Although a few dedicated designs for this regime have been discussed in the literature (e.g., Sakurai et al. 1990; Costa et al., 1993; Tomita et al. 1996; Cline et al. 1997; McConnell et al. 2002a; Curado da Silva et al. 2003; Larsson \& Pearce 2004), none have yet led to an operational instrument.

One dedicated design, referred to as GRAPE (the Gamma RAy Polarimeter Experiment), has been developed for the 50--300 keV energy range by McConnell et al. (1993, 1998, 1999a, 1999b, 2000, 2002a, 2002b; see also Legere et al. 2005).  The design is based on Compton scattering from a low-Z plastic scintillator into a high-Z inorganic scintillator (CsI or LaBr$_3$). Its very wide FoV also makes it ideal for studying the polarization of $\gamma$-ray bursts.  The GRAPE concept (Fig. 3) permits tiled arrays, to serve as either a large area, wide FoV detector or as a detection plane for a coded mask imaging polarimeter. A GRAPE science model has been demonstrated in the laboratory (Fig. 4).  Similar designs have been considered by Suzuki et al. (2003) and Produit et al. (2005).

In contrast to GRAPE, the Polarized Gamma-ray Observer (PoGO; Larsson \& Pearce 2004; Mizuno et al. 2005) is a collimated design which provides greater sensitivity for point sources in the 30-100 keV energy range.  Its limited FoV ($\sim3^{\circ}$) makes it ideal for known point sources, but not very effective for GRBs.  Curado da Silva et al. (2003) discuss an imaging polarimeter called CIPHER (Coded Imager and Polarimeter for High Energy Radiation) based on an array of CdTe detectors.   Using coded mask imaging, it would operate in the 100 keV -- 1 MeV energy range with a FoV of $\sim5^{\circ}$.

\begin{figure}
	\begin{minipage}[t]{.40\linewidth}
	\centering
	\includegraphics[width=2.50in]{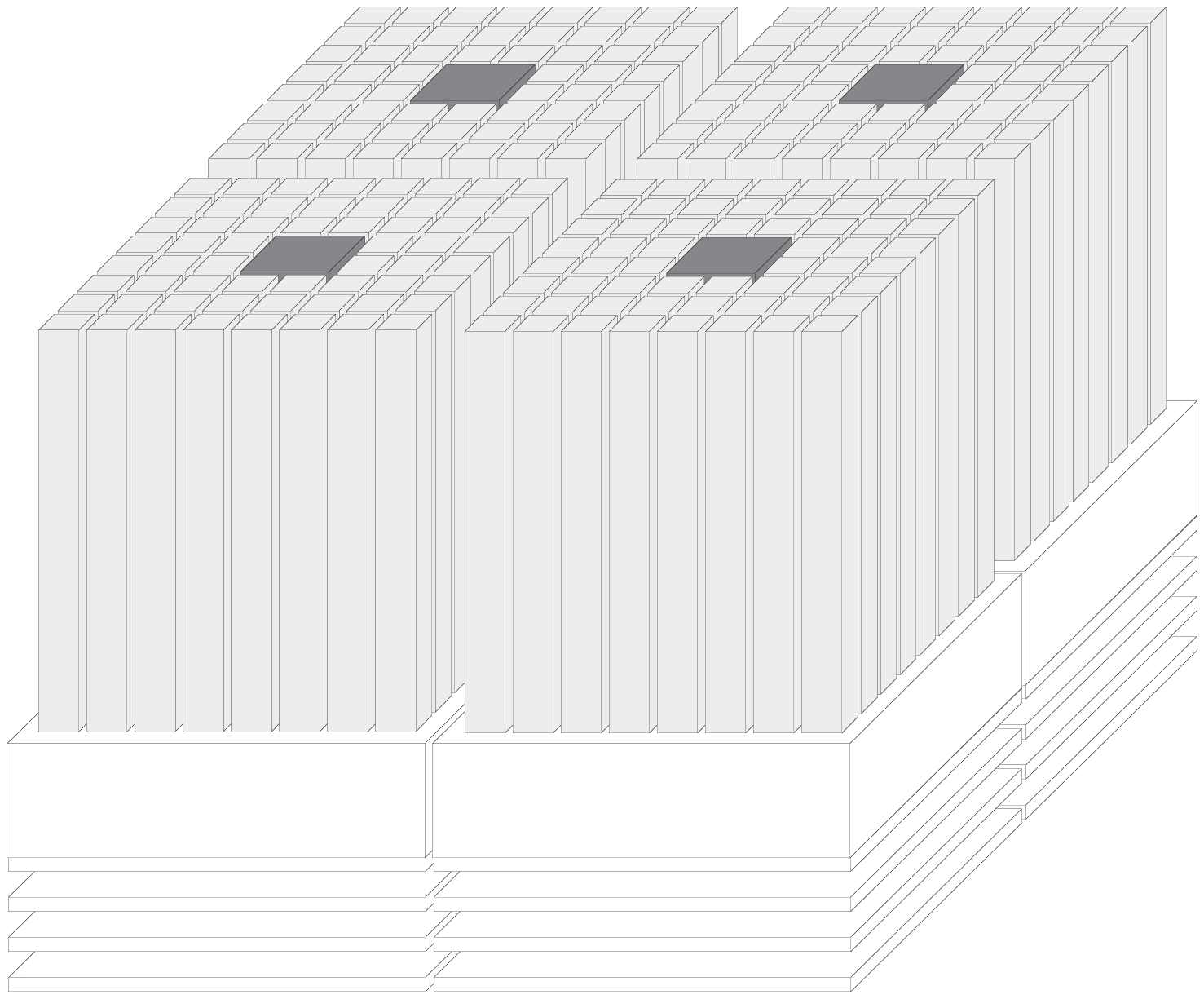}
	\caption{The GRAPE design utilizes a flat-panel MAPMT for readout, shown here in a tiled configuration.}
	\label{fig_sim}
	\end{minipage}\hfill
	\begin{minipage}[t]{.55\linewidth}
	\centering
	\includegraphics[width=3.00in]{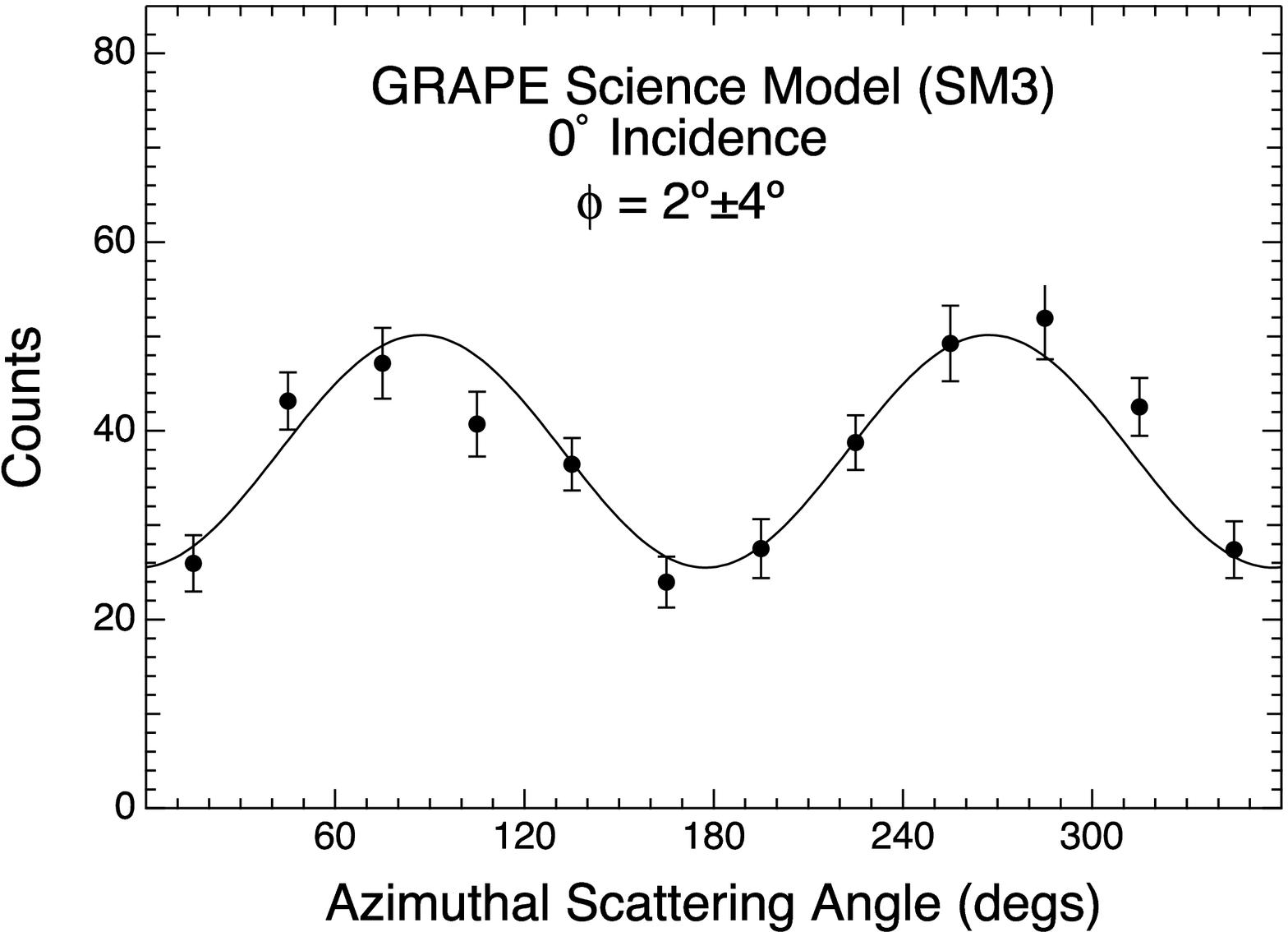}
	\caption{A measured result from laboratory tests of a single GRAPE module.  The measured polarization level of $\sim56$\% is consistent with that expected from the scattered beam of 662 keV photons used for the test.}
	\label{fig_sim}
	\end{minipage}
\end{figure}

\subsection{Medium Energy Gamma Rays (300 keV -- 30 MeV)}

At higher energies (300 keV -- 30 MeV), Compton polarimeters based on the use of high-Z scattering elements (coupled with high-Z absorbers) become viable.  For example, the Ge double scatter approach used by RHESSI becomes most effective at energies above $\sim300$ keV.  Multiple scatter events in high-Z coded mask detection planes also offer possibilities for polarimetry.  The use of a Ge strip detector has been demonstrated in this energy range (Kroeger et al. 1999).  An imaging polarimeter based on the use of CdTe is also being developed (Caroli et al. 2000).  In principle, both the IBIS and SPI instruments on INTEGRAL are capable of polarimetry in this energy band (Lei, Dean and Hills 1997; Stephen et al. 2001; Caroli et al. 2000).  Unfortunately, the lack of rotation makes the polarization analysis of these data difficult and telemetry limitations limit the capabilities of IBIS.  Efforts to measure GRB polarization with INTEGRAL have so far been unsuccessful.  In principle, the CdZnTe detection plane of the Swift BAT instrument (Barthelmy 2000) might make for a good polarimeter, but the packaging design of the detectors and the design of the signal processing electronics results in a loss of the necessary multiple scatter event information.  

This energy range is also the domain of Compton telescopes.  A properly configured Compton telescope can serve as a very powerful polarimeter.  The one Compton telescope that has flown in orbit, the COMPTEL instrument on CGRO  (Sch{\"o}nfelder et al. 19993), was very limited in its ability to do polarimetry.  This was due both to its inability to precisely measure the interaction sites and also to a very poor Compton scattering geometry that required scatter angles $<90^{\circ}$.  Although some efforts have been made to study polarization with COMPTEL data, no successful results have so far been obtained (Lei et al. 1996).

Compton telescope designs that are currently being studied offer a much more favorable geometry for polarization measurements.  With the elimination of time-of-flight measurements, recent designs are much more compact.  This results in significantly improved detection efficiency and a significantly larger FoV.  It also provides a far more optimized well-type geometry for Compton polarimetry.  The next generation of Compton telescopes will offer substantial improvements in polarization sensitivity.  Recent Compton telescope designs can be characterized as those that attempt to track the scattered electron, such as TIGRE (O'Neill et al.  1996) and MEGA (Kanbach et al. 2001), and those that don't, such as LXeGRIT (Aprile et al. 1984) and NCT (Boggs et al. 2001).  One concept for the Advanced Compton Telescope (ACT) involves a large (1 m$^2$) stack of Si strip detectors that is used to track multiple Compton intertactions (Kurfess \& Kroeger 2001).

\subsection{High Energy Gamma Rays ($>30$ MeV)}

The potential utility of pair production for measuring polarization at energies above 2 MeV has been recognized for some time (Maximon \& Olsen 1962).  Unfortunately, effective polarization measurements with pair production telescopes are limited by the effects of multiple coulomb scattering, which makes it difficult to define the plane of pair production.  Efforts to measure polarization both with COS-B (Mattox et al. 1990) and with CGRO/EGRET (Mattox 1991) have been unsuccessful, largely for this reason.  It also appears that both GLAST and AGILE will suffer from similar difficulties, making polarization measurements with those instruments unlikely. One recent design for an effective pair production polarimeter involves the use of gas micro-well detectors for tracking the electron-positron pair with minimal scattering (Bloser et al. 2003).

\subsection{Albedo Polarimetry}

A measurement of the linear polarization of a transient event (such as
a $\gamma$-ray burst or solar flare) can, in principle, be made by measuring the
angular distribution of the albedo flux, i.e., the source flux which
is scattered from the Earth's atmosphere prior to reaching the
detector.  This concept is based on the properties of the Compton
scattering of polarized radiation.  In particular, this approach
relies on the fact that, in the case of linearly polarized radiation,
the scattered photon tends to be ejected at right angles to the
electric field vector of the incident radiation.  The
atmosphere, as seen from an orbiting satellite, presents a wide range
of possible scatter angles for a given source direction.  The photon
scatter angle will depend on look direction.  Hence, the intensity
distribution of the albedo flux will exhibit an angular distribution
which will depend on the polarization properties of the 
source radiation.

This technique was originally explored by McConnell et al. (1996) using data from the BATSE experiment on CGRO.  More recently, the albedo polarimetry concept was used by Wills et al. (2005) to place constraints on GRB  polarization.  This approach is limited in its ability to measure polarization as a function of energy (since information on the original energy spectrum is lost through the scattering process), but it does offer the possibility to monitor a large fraction of the full sky.  The BATSE experiment was limited in its ability to image the albedo flux.  To take full advantage of this concept would require an instrument designed specifically for the task.

\section{Summary}

We can expect that, within the next 5--10 years, the field of $\gamma$-ray polarimetry will be returning useful information regarding various high energy phenomena in the universe.  The first experimental steps are now being taken and it is clear that the scientific return of such an endeavor will be of great interest.


\begin{acknowledgements}
This work has been partially supported by NASA grants NAG5-10203 and NAG5-5324.
\end{acknowledgements}


\bibliographystyle{aipprocl} 

\bigskip
\noindent
{\b DISCUSSION}
\bigskip

\noindent
{\b MASSIMO CAPPI:} Could you comment also on current possibilities to study polarimetry of soft X-rays (below 10 keV)?
\bigskip

\noindent
{BLOSER:} There are several groups pursuing this as well and their progress is also very exciting.  Two groups, at GSFC and in Italy, are using gas detectors with very fine spatial resolution to track the photoelectron that is ejected after a photoelectric absorption of the X-ray.  The azimuthal direction of the photoelectron is also strongly correlated with the incident polarization vector.  Other groups in the USA, at MSFC and at CFA, are using more "classical" methods of X-ray scattering in crystals.
\bigskip

\noindent
{\b ALDO MORSELLI:} What is the status and time schedule of the various experiments on polarization and which is the first that will fly?
\bigskip

\noindent
{BLOSER:} I am optimistic that at least some of the balloon experiments will be able to make useful measurements in the next several ($\sim5$?) years.  Both GRAPE and PoGO plan to fly test payloads within the next year or two.  NCT has recently flown its first engineering test flight.  The schedule for new satellite instruments is far less certain, mainly due to funding availability.  For example, we plan to propose MEGA as a MIDEX mission at the next opportunity -- but it is unclear as to when the next MIDEX opportunity will arise.
\bigskip

\label{lastpage}

\end{document}